\documentclass[12pt]{article}
\usepackage{graphicx,amssymb,amsmath,amsthm,amsfonts,epsfig}
\usepackage{color}
\usepackage{graphicx}
\begin{document}

\begin{center}
{\large \bf Entropy Product Formula for Gravitational Instanton}
\end{center}

\vskip 5mm

\begin{center}
{\Large{Parthapratim Pradhan\footnote{E-mail: pppradhan77@gmail.com}}}
\end{center}

\vskip  0.5 cm

{\centerline{\it Department of Physics}}
{\centerline{\it Hiralal Mazumdar Memorial College For Women}}
{\centerline{\it Dakshineswar, Kolkata-700035,  India.}}
\vskip 1cm

\begin{abstract}
We investigate the entropy product formula for various gravitational instantons. We speculate that due to the mass-independent 
features of the said instantons they are \emph{universal} as well as they are \emph{quantized}. For isolated 
Euclidean Schwarzschild black hole, these properties simply \emph{fail}. 
\end{abstract}


\section{Introduction}
There has been a strong interest in  microscopic interpretation of black hole (BH) entropy 
\cite{bk,bk1,bk2,bc,bcw,hk,hk1,hk3} in terms of $D$-branes come due to the work by Strominger and Vapa \cite{vapa96}. 
In $d$-dimension Euclidean quantum gravity, this entropy is due to the  $(d-2)$-dimensional fixed point sets
of the imaginary time translation Killing vector. There are many fixed point sets which can
also give rise to BH entropy.

Previously, Area (or Entropy) product formula evaluated for different class of BHs 
\cite{larsen,ansorg1,cvetic,castro,visser,det,chen,interior,pp14,hl,jetpl,abg,5d,ahep,extend,loga,sen,jetp,epl}.
In some cases, the product formula is not mass-independent (universal) and in some cases the product formula is 
indeed mass-independent i.e. universal. Upto the author's knowledge, there has been no attempt to compute 
the entropy product formula for \emph{gravitational instanton}.

Thus in the present work, we wish to investigate the entropy product formula 
for various gravitational instantons. Instantons are non-singular and having imaginary time.
They arises in quantum field thory (QFT) for evaluating the functional integral, in which 
the functional integral is Wick rotated and expressed as an integral over Euclidean field configuarations. They
are the solutions of the Euclidean Einstein equations. They have the signature of the form $(++++)$.
There are two types of instantons discoverd so far. One is an asymptotically locally flat (ALF),
which was first discoverd by Hawking in 1977 \cite{haw77} and the other is an asymptotically
locally Euclidean (ALE), discovered by Gibbons and Hawking in 1978 \cite{gh79}. The examples
of ALE classes are Flat space, Eguchi-Hanson \cite{eh78} and multi-instanton \cite{gh79}.

The ALF class of solutions are asymptotically flat (AF) in 3D sense, the fourth, imaginary
-time, direction being periodic. The surfaces of large radii should be think of as an
${\cal S}^1$ bundle over ${\cal S}^2$. The product bundle corresponds to the AF solutions which include 
the Euclidean Schwarzschild and Euclidean Kerr solutions \cite{gh77,gp80}. The twisted bundles correspond 
to the multi-Taub-NUT solution \cite{haw77}, and the Taub-Bolt solution was first discoverd 
by Page \cite{page78}.

In our previous work \cite{jetp,epl,pp14}, we investigated the properties of inner and outer horizon
thermodynamics of Taub-NUT (Newman-Unti-Tamburino) BH, Kerr-Taub-NUT BH and Kerr-Newman-Taub-NUT BH in
four dimensional Lorentzian geometry. The failure of \emph{First law} of BH
thermodynamics and \emph {Smarr-Gibbs-Duhem  relation for Taub-NUT and Kerr-Taub-NUT
BH in the Lorentzian regime gives the motivation behind this work}. What happens
when one can go from \emph{Lorentzian geometry to Riemannian geometry?} This is the
prime aim in this work. By studing the properties of these instantons what should be 
the effects on the BH entropy product formula  due to the non-trivial NUT parameter?

In general relativity, the non-trivial value for the BH entropy is due to the
presence of the fixed point set of the periodic imaginary time Killing vector. The
fixed point set here we considered actually is the BH horizons (${\mathcal H}^{\pm}$).
Here ${\mathcal H}^{+}$ is called event horizon and ${\mathcal H}^{-}$ is called
the Cauchy horizon. In four dimension, such fixed point sets are of two types, isolated
points or zero dimensional which we call NUTs and two surfaces or two dimensional which
we call Bolts. Thus one can thought Bolts as being the analogue of electric type mass-
monopoles and the NUTs as being gravitational dyons endowed with a real electric type
mass-monopole and an imaginary magnetic type mass-monopole. The presence of magnetic
type mass introduces a Dirac string like singularity in the spacetime which is so called
Misner string was first pointed out by Misner in his paper for the Lorentzian Taub-NUT
spacetime \cite{mis63}. A Misner string is a coordinate singularity which can be considered
as a manifestation of a ``non-trivial topological twisting'' \cite{aliev} of the manifold 
($M,g_{ab}$). This twist is parametrized by a topological term, the NUT charge.

In our previous investigation \cite{pp14}, we have taken the metric in a Lorentzian spacetime in
$3+1$ split form as
\begin{eqnarray}
ds^2 &=& - {\cal F} \, \left(dt+w_{i}dx^{i}\right)^2+ \frac{\gamma_{ij}}{{\cal F}}dx^{i}dx^{j}
~.\label{lor}
\end{eqnarray}

In this work, we are interested to study the metric in a \emph{Riemannian} spacetime which can
be written in $3+1$ split form:
\begin{eqnarray}
ds^2 &=& {\cal F} \, \left(d\tau+w_{i}dx^{i}\right)^2+ \frac{\gamma_{ij}}{{\cal F}}dx^{i}dx^{j}
~.\label{rie}
\end{eqnarray}
Here all quantities are independent of $t$ or $\tau$. The Wick rotation that transforms from
one case to other is by the transformation  $t\mapsto i\tau$ and $w_{i}\mapsto i\omega_{i}$.
${\cal F}$ can be thought of as an electric type potential, and $\omega_{i}$ or $w_{i}$ as a
magnetic-type vector potential. The associated magnetic field is
$H_{ij}=\partial_{i}\omega_{j}-\partial_{j}\omega_{i}$ and it is gauge invariant.

This should be used to define a magnetic  monopole moment called the NUT 
charge $n$ \cite{gh79}. If $n\neq 0$, the fibration should not be trivial. 
In the Lorentzian geometry, these fixed point sets are the two-dimensional Boyer bifurcation 
sets of event horizon \cite{bcw,boy68}. On the other hand in Riemannian geometry, these 
fixed point sets are of two types: zero dimensional point or NUTs, and two dimensional surfaces or Bolts \cite{gh79}. 
A NUT possesses a pair of surface gravities $\kappa_{1}$ and $\kappa_{2}$. $p$ and $q$ are a pair of co-prime integers 
such that $\frac{\kappa_{1}}{\kappa_{2}}=\frac{p}{q}$. If $\frac{\kappa_{1}}{\kappa_{2}}$ is
irrational, $p=q=1$. A NUT of type $(p,q)$ has a NUT charge of $n=\frac{\beta}{8 \pi pq}$, where $\beta$ is the 
period of the imaginary time coordinate. Moreover, $n=\frac{Y \beta}{8\pi}$ for a Bolt of 
self-intersection number $Y$. It should be noted that in the Riemannian case, the number of NUTs and Bolts are 
related to the Euler number $\chi$ and the Hirzebruch signature $\tau$ of the manifold $M$ by
\begin{eqnarray}
 \chi &=& \sum_{Bolts} \chi_{i}+\sum_{NUTs} 1, 
\end{eqnarray}
where $\chi_{i}$ is the Euler number for the $i-$th Bolt, and 
\begin{eqnarray}
\tau &=& \sum_{Bolts} Y_{i} \csc^{2} \theta -\sum_{NUTs} \cot p_{i}\theta \cot q_{i}\theta+\eta(0, \theta) ~.\label{tau1}
\end{eqnarray}
The above Eq. (\ref{tau1}) is valid for arbitrary $\theta$. $Y_{i}$ is the self-interection number of the 
$i$-th Bolt, the $i-$th NUT is of type $(p_{i}, q_{i})$ and $ \eta(0,  \theta)$ is a correction term which 
depends solely on the boundary. It should be mentioned that for multi-Taub-NUT and multi-instanton solutions there are 
$k$ NUTs of type $(1, 1)$ and $\tau=k-1$, therefore 
\begin{eqnarray}
\tau &=& \sum_{Bolts} Y_{i} \csc^{2} \theta -\sum_{NUTs} \cot p_{i}\theta \cot q_{i}\theta+k\csc^2\theta-1 
~.\label{tau2}
\end{eqnarray}
where the AF boundary conditions should be $\eta(0, \theta)=0$ \cite{gp80}.

The structure of the paper is as follows. In Sec. \ref{eusc}, we have considered the Euclidean Schwarzschild 
BH. In Sec. \ref{etn}, we have investigated the properties of Self dual Taub-NUT instantons. In Sec. \ref{etn1}, 
we have studied the mass-independent properties of Taub-BoltiInstantons. In Sec. \ref{eh}, we have described the 
properties of  Eguchi-Hanson instantons. In Sec. \ref{adtn}, we have examined the properties of  Taub-NUT-AdS Spacetime.
In Sec. \ref{adtb}, we have studied the entropy product formula for Taub-Bolt-AdS spacetime and finally in Sec. \ref {dadtn}, 
we have examined the product rules for Dyonic Taub-NUT-AdS and Taub-Bolt-AdS spacetime.

\section{\label{eusc} Euclidean Schwarzschild metric:}
To give a warm up, let us first consider the Schwarzschild BH (where we have used units in $G=c=1$) in Euclidean form as
\begin{eqnarray}
ds^2 &=& \left(1-\frac{2M}{r}\right) d{\tau}^2+ \frac{dr^2}{\left(1-\frac{2M}{r}\right)}
+r^2 \left(d\theta^2+\sin^2\theta d\phi^2 \right) ~.\label{esch}
\end{eqnarray}
The apperent singularity at the event horizon $r_{+}=2M$ can be removed by identifying $\tau$
with a period $\beta=8 \pi M$ \cite{gp80,haw77}. The radial coordinate has the range
$2M \leq r \leq \infty$. Then the topology of the manifold is $R^2\times {\cal S}^2$. The
isometry group is $O(2)\otimes O(3)$, where the  $O(2)$ corresponds to translations in the
periodically identified imaginary time $\tau$ and the  $O(3)$ corresponds to rotations of the
$\theta$ and $\phi$ coordinates.

The Killing vector $\partial_{\tau}$ has unit magnitude at large radius and has a Bolt on the
horizon $r_{+}=2M$ which is a 2-sphere ${\cal S}^2$ of area
\begin{eqnarray}
A_{+} &=& 16\pi M^2 ~.\label{esar}
\end{eqnarray}
The surface gravity is given by
\begin{eqnarray}
\kappa_{+} &=& \frac{2 \pi}{\beta}= \frac{1}{4M}~.\label{kesh}
\end{eqnarray}
and the BH temperature is
\begin{eqnarray}
T_{+} &=& \frac{\kappa_{+}}{2\pi} =\frac{1}{8\pi M} ~.\label{tmesh}
\end{eqnarray}
Thus for an isolated Euclidean Schwarzschild BH the area product becomes
\begin{eqnarray}
A_{+} &=& 16\pi M^2 ~.\label{esar1}
\end{eqnarray}
which tells us that the product is dependent on mass parameter and thus it is not universal. Also it 
is not quantized.

The Euclidean action derived in \cite{haw77}
\begin{eqnarray}
I= -\ln Z=4 \pi M^2 ~.\label{acesh}\\
\end{eqnarray}
From that one can derive the entropy as in \cite{haw77}
\begin{eqnarray}
S_{+} = -\left(\beta \frac{\partial}{\partial \beta}-1\right)\ln Z=4 \pi M^2 ~.\label{sesh}
\end{eqnarray}
Thus the entropy product for isolated Euclidean Schwarzschild BH should be
\begin{eqnarray}
S_{+} =4 \pi M^2 ~.\label{sesh1}
\end{eqnarray}
Indeed, it is not universal as well as  it is not quantized.

\section{\label{etn} Self dual Taub-NUT Instantons:}
In this section we shall calculate the entropy product and area product of four dimensional
Taub-NUT spacetime. It is of ALF. ALF metrics have a NUT charge,or magnetic type mass, $n$,
as well as the ordinary electric type mass, $M$. The NUT charge is $\frac{\beta c_{1}}{8\pi}$,
where $c_{1}$ is the first Chern number of the $U(1)$ bundle over the sphere at infinity, in
the orbit space $\Xi$. If $c_{1}=0$, then the boundary at infinity is
${\cal S}^1\times {\cal S}^2$ and the spacetime is AF. The BH metrics are saddle
points in the path integral for the partition function. Thus, if $c_{1}\neq 0$, the boundary
at infinity is a squashed ${\cal S}^3$, and the metric should not be analytically 
continued to a Lorentzian metric. The squashed ${\cal S}^3$ is the three-dimensional space on which
the boundary conformal field theory (CFT) will be compactified, with $\beta$ identified with
the inverse temperature i. e. $T=\frac{1}{\beta}$.

Hawking \cite{haw77} first given an examples of gravitational instanton was the self-dual
Taub-NUT metric described by
\begin{eqnarray}
ds^2 &=& {\cal F}(r) \, \left(d\tau+2n\cos\theta d\phi\right)^2+ \frac{dr^2}{{\cal F}(r)}
+\left(r^2-n^2\right) \left(d\theta^2+\sin^2\theta d\phi^2 \right) ~.\label{tnn}\\
{\cal F}(r) &=& \frac{r-n}{r+n}
\end{eqnarray}
It is ALF with a central NUT. The self-dual Taub-NUT instanton has $M=n$ and
the anti-self-dual instanton has $M=-n$. The value $r=n$ is now a zero in ${\cal F}(r)$.
The $(\theta, \phi)$ two-sphere has a zero area at $r=n$, so the zero in ${\cal F}(r)$
is a zero-dimensional fixed point of $\partial_{\phi}$, a NUT.

In order to make the solution regular, we take the region $r\geq n$ and let the period
of $\tau$ be $8\pi n$. The metric has a NUT at $r=n$, with a Misner string running along
the $z$-axis from the NUT out to infinity i.e. $n\leq r \leq \infty$.

We know from the idea of path-integral formulation of quantum gravity which tells us that
the Euclidean action derived in \cite{haw77}
\begin{eqnarray}
I &=& -\ln Z=4 \pi n^2 ~.\label{act}
\end{eqnarray}
where $Z$ is the partition function of an ensemble
\begin{eqnarray}
Z &=& \int [Dg] [D\phi] e^{-I(g,\phi)} ~.\label{zact}
\end{eqnarray}
with the path integral taken over all metrics $g$ and matter field $\phi$ that are
appropriately identified with the period $\beta$ of $\tau$.
Therefore the entropy should be derived as
\begin{eqnarray}
S &=& -\left(\beta \frac{\partial}{\partial \beta}-1\right)\ln Z=4 \pi n^2 ~.\label{sea}
\end{eqnarray}
It is indeed mass-independent and thus it is universal.

The surface gravity is calculated to be
\begin{eqnarray}
{\kappa} &=& \frac{2 \pi}{\beta}=\frac{1}{4n}~.\label{ket}
\end{eqnarray}
Thus the BH temperature should read off
\begin{eqnarray}
T &=&\frac{\kappa}{2\pi} =\frac{1}{8\pi n} ~.\label{tmtn}
\end{eqnarray}

Now we see what happens the above results for another instantons that is Taub-Bolt.

\section{\label{etn1} Taub-Bolt Instantons:}
The Taub-Bolt instanton is described by the metric \cite{page78}

\begin{eqnarray}
ds^2 &=& {\cal G}(r) \, \left(d\tau+2n\cos\theta d\phi\right)^2+ \frac{dr^2}{{\cal G}(r)}
+\left(r^2-n^2\right) \left(d\theta^2+\sin^2\theta d\phi^2 \right) ~.\label{tbe}\\
{\cal G}(r) &=& \frac{(r-2n)(r-\frac{n}{2})}{(r^2-n^2)}=\frac{(r-r_{+})(r-r_{-})}{(r+n)(r-n)}
\end{eqnarray}

It is a non-self-dual, non-compact solution of the vacuum Euclidean Einstein equations. In order
to make the solution regular we have restricted in the region $r=r_{+}\geq 2n$, and the Euclidean time
has period $\beta=8 \pi n$. Asymptotically, the Taub-Bolt instanton behaves similar manner as
the Taub-NUT, so it is ALF. Since we are setting the fixed point is at $r=r_{+}=2n$, therefore the
area of the ${\cal S}^2$ does not vanish there and the fixed point set is 2-dimensional,
thus it is a Bolt of area
\begin{eqnarray}
A_{+} &=& 12 \pi n^2 ~.\label{ebt}
\end{eqnarray}
Thus the area product for Taub-Bolt instanton will be
\begin{eqnarray}
A_{+} &=& 12 \pi n^2 ~.\label{ebt1}
\end{eqnarray}
Thus the area product does independent of mass and also quantized.
Similarly, the action was calculated in \cite{hh99}
\begin{eqnarray}
I= -\ln Z= \pi n^2 ~.\label{actb1}
\end{eqnarray}
Thus the entropy was derived by the universal formula
\begin{eqnarray}
S_{+} = -\left(\beta \frac{\partial}{\partial \beta}-1\right)\ln Z= \pi n^2 ~.\label{stb}
\end{eqnarray}
It indicates that the entropy product should be universal and quantized.

\section{\label{eh} Eguchi-Hanson Instantons:}
A non-compact instanton which is a limiting case of the Taub-NUT solution is the
Eguchi-Hanson metric \cite{eh78},
$$
ds^2=\left(1-\frac{n^4}{r^4}\right) \left(\frac{r}{8n}\right)^2 \left(d\tau+4n\cos\theta d\phi\right)^2
+ \frac{dr^2}{\left(1-\frac{n^4}{r^4}\right)}
$$
\begin{eqnarray}
+\frac{r^2}{4}\left(d\theta^2+\sin^2\theta d\phi^2 \right) ~.\label{ehmt}
\end{eqnarray}
The instanton is regular if we consider the region $r\geq n$, and let $\tau$
has period $8\pi n$. The metric is ALE type. There is a Bolt of area at $r=n$
is given by
\begin{eqnarray}
A &=& \pi n^2 ~.\label{ehar}
\end{eqnarray}
which gives rise to a Misner string along the $z$-axis. Thus the product is universal and should
be quantized.
The Euclidean action derived in \cite{hh99}
\begin{eqnarray}
I= 0 ~.\label{aceh}
\end{eqnarray}
Thus entropy corresponds to
\begin{eqnarray}
S = \left(\beta \frac{\partial}{\partial \beta}-1\right)I=0 ~.\label{seh}
\end{eqnarray}

We now turn our attention for the Taub-NUT and Taub-Bolt geometries in four dimensional
locally AdS spacetime. The spacetimes have a global non-trivial topology
due to the fact that one of the Killing vector has a zero dimensional fixed point set
called NUT or a two-dimensional fixed point set called Bolt. Moreover, these four
dimensional spacetimes have have Euclidean sections which can not be exactly matched
to AdS spacetime at infinity.

\section{\label{adtn} Taub-NUT-AdS Spacetime:}
In this section we shall consider the spacetime which are only locally asymptotically
AdS and with non-trivial topology. The metric on the Euclidean section of this family
of solutions could be written as \cite{pope86,pope87}

\begin{eqnarray}
ds^2 &=& {\cal H}(r) \, \left(d\tau+2n\cos\theta d\phi\right)^2+ \frac{dr^2}{{\cal H}(r)}
+\left(r^2-n^2\right) \left(d\theta^2+\sin^2\theta d\phi^2 \right) ~.\label{tnads}
\end{eqnarray}
where
\begin{eqnarray}
{\cal H}(r) &=& \frac{\left(r^2+n^2\right)-2Mr+\ell^{-2}\left(r^4-6n^2r^2-3n^4\right)}{r^2-n^2}
\end{eqnarray}
and  $\ell^2=\frac{-3}{\Lambda}$, with $\Lambda <0$ being the cosmological constant.
Here $M$ is a (generalized) mass parameter and $r$ is a radial
coordinate. Also, $\tau$, the analytically continued time i.e. Euclidean time,
parametrizes a circle ${\cal S}^1$, which is fibered over the two-sphere ${\cal S}^2$, with
coordinates $\theta,\phi$. The non-trivial fibration is a consequence of a non-vanishing
NUT parameter $n$.

There are some restrictions \cite{myers99} for existence of a regular NUT parameter. Firstly, in order to
ensure that the fixed point set is zero dimensional, it is necessary that the Killing vector
$\partial _{\tau}$ has a fixed point which occurs precisely when the area of the two-sphere
is zero size.
Secondly, in order for the Dirac-Misner string \cite{mis63} to be unobservable, it is necessary that
the period of $\tau$ be $\beta=8\pi n$. To avoid the conical singularity, we must check
${\cal H}'(r_{+}=n)=\frac{1}{2n}$.
Thirdly, the mass parameter $M$ must be $M=n-\frac{4n^3}{\ell^2}$.
After simplifying the metric coefficients, we obtain
\begin{eqnarray}
{\cal H}(r) &=& \left(\frac{r-n}{r+n}\right)\left[1+\frac{(r-n)(r+3n)}{\ell^2}\right]
\end{eqnarray}
and the range of the radial coordinate becomes $n\leq r\leq \infty$.
For our requirement, the Euclidean action for this spacetime was
calculated in \cite{mann99,empa99}
\begin{eqnarray}
I= -\ln Z= 4\pi n^2\left(1-\frac{2n^2}{\ell^2}\right) ~.\label{actnd}
\end{eqnarray}
and
the entropy will be
\begin{eqnarray}
S_{+} &=& \left(\beta \frac{\partial}{\partial \beta}-1\right)I=
4\pi n^2\left(1-\frac{6n^2}{\ell^2}\right) ~.\label{snads}
\end{eqnarray}
Thus the entropy product should be
\begin{eqnarray}
S_{+} &=& 4\pi n^2\left(1-\frac{6n^2}{\ell^2}\right) ~.\label{snads1}
\end{eqnarray}
It is independence of mass parameter and does depend on NUT parameter and
cosmological constant. Thus the entropy product is universal for Taub-NUT-AdS
spacetime. 
The Hawking temperature $T_{+}=\frac{1}{8 \pi n}$ is
same as Taub-NUT BH. The first law of thermodynamics is also satisfied as
$dM=T_{+}dS$. 

If we consider the \emph{extended phase space} following our previous work \cite{extend} and 
in this framework, the cosmological constant ($\Lambda$) should be treated as thermodynamical pressure i.e. 
$P = -\frac{\Lambda}{8\pi}=\frac{3}{8\pi \ell^2} $ and 
its conjugate variable should be treated as thermodynamic volume i.e. 
$V_{+} = \frac{4}{3}\pi r_{+}^3$, where $r_{+}$ is the 
horizon radius. Then one should be interpreted the ADM mass $M$ parameter not to be the energy rather it should 
be interpreted as enthalpy $H=M=U+PV_{+}$ of the gravitational thermodynamical system. Therefore the 
thermodynamic volume has been calculated in \cite{cv14} for Taub-NUT-AdS spacetime:
\begin{eqnarray}
V_{+} &=& \left(\frac{\partial H}{\partial P}\right)_{S}=-\frac{8}{3}\pi n^3  ~.\label{vmn}
\end{eqnarray} 
One aspect, this is a peculiar result in a sense that the thermodynamic volume is negative and the other aspect is that the 
thermodynamic volume is \emph{universal} because it is independent of the ADM mass parameter. It should be noted that the 
first law is fulfilled in this case and it yields
\begin{eqnarray}
dH &=&  T_{+} d{\cal S} + V_{+} dP ~. \label{eq9n}
\end{eqnarray}
Analogously, the Smarr-Gibbs-Duhem relation should be
\begin{eqnarray}
H &=&  2T_{+} {\cal S} - 2P V_{+} ~. \label{enq10}
\end{eqnarray}
and another interesting result we first claimed that the internal energy for Taub-NUT-AdS BH is \emph{universal}. It is 
given by 
\begin{eqnarray}
U &=& n \left(1-8\pi P n^2 \right) ~. \label{un}
\end{eqnarray}

\section{\label{adtb} Taub-Bolt-AdS Spacetime:}
For Taub-Bolt-AdS, the metric has the same form as in (\ref{tnads}) but the
fixed point set here is two dimensional or Bolt and  with additional restrictions are
the metric coefficients ${\cal H}(r)$ vanish at $r=r_{b}>n$. In order to have a regular
Bolt at $r=r_{b}$, the following conditions must be satisfied: (i) ${\cal H}(r_{b})=0$,
(ii) ${\cal H}'(r_{b})=\frac{1}{2n}$ and the numerator of ${\cal H}(r)$ at $r=r_{b}$
being a single one. From the condition (i), we get the mass parameter at $r=r_{b}$:
\begin{eqnarray}
 M &=& M_{b}=\frac{r_{b}^2+n^2}{2r_{b}}+\frac{1}{2\ell^2}\left(r_{b}^3-6n^2r_{b}-
3\frac{n^4}{r_{b}}\right) ~.\label{mads1}
\end{eqnarray}
Then we find\cite{myers99}
\begin{eqnarray}
 {\cal H}'(r_{b}) &=& \frac{3}{\ell^2} \left(\frac{r_{b}^2-n^2+\ell^2/3}{r_{b}}\right)
\end{eqnarray}
To satisfy the condition (ii) we must have the quadratic equation for $r_{b}$:
\begin{eqnarray}
6nr_{b}^2-\ell^2 r_{b}-6n^3+2n\ell^2 &=& 0 ~.\label{mads2}
\end{eqnarray}
which gives the solution for $r_{b}$ in two branches
\begin{eqnarray}
r_{b\pm} &=& \frac{\ell^2}{12n}\left(1\pm \sqrt{1-48\frac{n^2}{\ell^2}+144 \frac{n^4}{\ell^4}} \right)
~.\label{mads3}
\end{eqnarray}
The discriminat of the above equation must be negative for $r_{b}$ to be real and for
$r_{b}>n$ we obtain the following inequality for $n$:
\begin{eqnarray}
 n \leq n_{max}=\sqrt{\frac{1}{6}-\frac{\sqrt{3}}{12}}\ell ~.\label{rann}
\end{eqnarray}
The Euclidean action was computed in \cite{empa99}
\begin{eqnarray}
I &=& \frac{4\pi n}{\ell^2}\left(M_{b}\ell^2+3n^2r_{b}-r_{b}^3 \right) ~.\label{aadsb}
\end{eqnarray}
Now it can be easily derive the entropy via the universal entropy formula:
\begin{eqnarray}
S_{+} &=& \left(\beta \frac{\partial}{\partial \beta}-1\right)I=
4\pi n \left(M_{b}-\frac{3n^2r_{b}}{\ell^2}+\frac{r_{b}^3}{\ell^2}\right) ~.\label{sbads}
\end{eqnarray}
Now substituating the values of $M_{b}$, we find the value of entropy
\begin{eqnarray}
S_{+} &=& 4\pi n \left[\frac{r_{b}^2+n^2}{2r_{b}}+\frac{1}{2\ell^2}\left(3r_{b}^3-12n^2r_{b}-
3\frac{n^4}{r_{b}}\right) \right] ~.\label{sbadsf}
\end{eqnarray}
Again putting the values of $r_{b}$, we see that the entropy is universal as well as quantized.

\section{\label{dadtn} Dyonic Taub-NUT-AdS and Taub-Bolt-AdS Spacetime:}
The general form of the metric for dyonic Taub-NUT-AdS spacetime \cite{pb75,pd76,or00} is given by
\begin{eqnarray}
ds^2 &=& {\cal N}(r) \, \left(d\tau+2n\cos\theta d\phi\right)^2+ \frac{dr^2}{{\cal N}(r)}
+\left(r^2-n^2\right) \left(d\theta^2+\sin^2\theta d\phi^2 \right) ~.\label{dtnads}
\end{eqnarray}
where,
\begin{eqnarray}
{\cal N}(r) &=& \frac{\left(r^2+n^2+4n^2\nu^2-q^2\right)-2Mr+\ell^{-2}\left(r^4-6n^2r^2-3n^4\right)}{r^2-n^2}
\end{eqnarray}
The gauge field reads off
\begin{eqnarray}
 A \equiv A_{\mu} dx^{\mu}=\left(\frac{qr}{r^2-n^2}+\nu \frac{r^2+n^2}{r^2-n^2}\right)
 \left(d\tau-2n \cos\theta d\phi \right),
\end{eqnarray}
The conditions of smoothness of the Euclidean section implies that the parameter $q$ is related
to the parameter $\nu$ gives a deformation from the uncharged system. When these parameters go
to to zero value, we obtain simply Taub-NUT-AdS spacetime.

In order to have a regular position of NUT or Bolt at $r=r_{\pm}$, we set ${\cal N}(r)=0$ and also
the gauge field $A$ must be regular at that point. Thus we obtain the mass parameter as
\begin{eqnarray}
M &=& \frac{r_{\pm}^2+n^2+4n^2\nu^2-\nu^2}{2r_{\pm}}+\frac{1}{2\ell^2}\left(r_{\pm}^3-6n^2r_{\pm}-
3\frac{n^4}{r_{\pm}}\right) ~.\label{mdtn}
\end{eqnarray}
and
\begin{eqnarray}
 q &=& -\frac{r_{\pm}^2+n^2}{r_{\pm}}\nu ~.\label{qdtn}
\end{eqnarray}
The electric charge and potential at infinity corresponds to

\begin{eqnarray}
 Q=q\,\, \phi_{\pm}= -q\frac{r_{\pm}}{r_{\pm}^2+n^2}=\nu ~.\label{qdtn1}
\end{eqnarray}

Now the Euclidean action for the above spacetime was calculated
in \cite{md06}(in units where $G=c=1$)
\begin{eqnarray}
I_{\pm} &=& -2\pi \frac{\left[r_{\pm}^4-\ell^2r_{\pm}^2+n^2\left(3n^2-\ell^2\right)\right]r_{\pm}^2
-\left(r_{\pm}^4+4n^2r_{\pm}^2-n^4\right)\ell^2\nu^2}{\left(3r_{\pm}^2-3n^2+\ell^2\right)r_{\pm}^2
+\left(r_{\pm}^2-n^2\right)\ell^2\nu^2} ~.\label{adtn1}
\end{eqnarray}
The entropy was calculated as:
\begin{eqnarray}
S_{\pm} &=& 2\pi \frac{\left[3r_{\pm}^4+\left(\ell^2-12n^2\right)r_{\pm}^2+n^2\left(\ell^2-3n^2\right)\right]r_{\pm}^2
+\left(r_{\pm}^4+4n^2r_{\pm}^2-n^4\right)\ell^2\nu^2}{\left(3r_{\pm}^2-3n^2+\ell^2\right)r_{\pm}^2
+\left(r_{\pm}^2-n^2\right)\ell^2\nu^2} ~.\label{sdtn}
\end{eqnarray}
When we set $r_{\pm}=r_{n}=n$, we get a dyonic NUT spacetime. For this spacetime
the above calculations reduced to
\begin{eqnarray}
M=n-\frac{4n^3}{\ell^2}, \,\, Q=-2n\nu, \phi_{\pm}=\nu ~.\label{qdtn2}
\end{eqnarray}
and
\begin{eqnarray}
 I_{\pm} &=& 4\pi n^2\left(1-2\frac{n^2}{\ell^2}+2\nu^2 \right) ~.\label{idn3}
\end{eqnarray}
Finally, the entropy is given by

\begin{eqnarray}
S_{\pm} &=& 4\pi n^2\left(1-6\frac{n^2}{\ell^2}+2\nu^2\right) ~.\label{idn4}
\end{eqnarray}
Thus the entropy product formula for dyonic Taub-NUT is
\begin{eqnarray}
S_{+}S_{-} &=& S_{+}^2=S_{-}^2 =(4\pi n^2)^2\left(1-6\frac{n^2}{\ell^2}+2\nu^2\right)^2 ~.\label{pdn1}
\end{eqnarray}
The product formula is indeed independent of mass and depends on $n$, $\ell$ and $\nu$.

For a dyonic Bolt, we set $r_{\pm}=r_{b}$ and satisfies the fourth order equation for $r_{b}$:
\begin{eqnarray}
6nr_{b}^4-\ell^2 r_{b}^3+2n\left(\ell^2-3n^2+\ell^2\nu^2\right)r_{b}^2-2\ell^2\nu^2n^3 &=& 0 ~.\label{qdb}
\end{eqnarray}
Let $r_{b}=r_{b\pm}$ be the solution of the equation. Then we find the action as previously

\begin{eqnarray}
I_{\pm} &=& -2\pi \frac{\left[r_{b}^4-\ell^2r_{b}^2+n^2\left(3n^2-\ell^2\right)\right]r_{b}^2
-\left(r_{b}^4+4n^2r_{b}^2-n^4\right)\ell^2\nu^2}{\left(3r_{b}^2-3n^2+\ell^2\right)r_{b}^2
+\left(r_{b}^2-n^2\right)\ell^2\nu^2} ~.\label{adtb1}
\end{eqnarray}
Similarly the entropy is given by
\begin{eqnarray}
S_{\pm} &=& 2\pi \frac{\left[3r_{b}^4+\left(\ell^2-12n^2\right)r_{b}^2+n^2\left(\ell^2-3n^2\right)\right]r_{b}^2
+\left(r_{b}^4+4n^2r_{b}^2-n^4\right)\ell^2\nu^2}{\left(3r_{b}^2-3n^2+\ell^2\right)r_{b}^2
+\left(r_{b}^2-n^2\right)\ell^2\nu^2} ~.\label{sdtb}
\end{eqnarray}
After substituating the value of $r_{b\pm}$ in the entropy product formula, it seems that
the product is independent of mass and depends on $n,\ell, \nu$ for dyonic Taub-Bolt instanton.

\section{\label{dis} Conclusion:}

We have studied the mass-independent feature for various gravitational instantons. This universal feature gives 
us strong indication towards understanding the microscopic properties of BH entropy. It would be 
interesting if one considered the entropy product formula for other instantons like Multi-Taub NUT, Non-self dual 
Taub-NUT, $S^4$, $CP^2$, $S^2\times S^2$ and Twisted $S^2\times S^2$. We expect these instantons also gives us 
universal features.



\end{document}